\documentclass[preprint]{emulateapj} 

\usepackage{natbib}
\usepackage{hyperref}
\hypersetup{colorlinks,citecolor=Blue,linkcolor=Red,urlcolor=Blue}
\usepackage{amsmath}
\usepackage{empheq}
\usepackage[usenames,dvipsnames]{color}

\allowdisplaybreaks 
\hypersetup{colorlinks,citecolor=Blue,linkcolor=Red,urlcolor=Blue}
\usepackage[usenames,dvipsnames]{color}
\usepackage{amsmath,amsfonts,amssymb}
\usepackage{graphicx}
\usepackage{wasysym}
\usepackage{float}
\usepackage{url}
\usepackage{longtable}
\usepackage{rotfloat}

\newcommand{\appropto}{\mathrel{\vcenter{\offinterlineskip\halign{\hfil$##$\cr\propto\cr\noalign{\kern2pt}\sim\cr\noalign{\kern-2pt}}}}}
\usepackage{threeparttable}

\begin{document}

\shorttitle{Kepler-9~b transit relocation}

\title{Transiting Exoplanet Monitoring Project (TEMP). III. On the Relocation of the Kepler-9~b Transit}  

\author{Songhu Wang$^{1,10}$, Dong-Hong Wu$^{2}$, Brett C. Addison$^{3}$, Gregory Laughlin$^{1}$, Hui-Gen Liu$^{2}$, Yong-Hao Wang$^{4,8}$, Taozhi Yang$^{5}$, Ming Yang$^{2}$, Abudusaimaitijiang Yisikandeer$^{5}$, Renquan Hong$^{6}$, Bin Li$^{6}$, Jinzhong Liu$^{5}$, Haibin Zhao$^{6}$, Zhen-Yu Wu$^{4,8}$, Shao-Ming Hu$^{7}$, Xu Zhou$^{4,8}$, Ji-Lin Zhou$^{2}$, Hui Zhang$^{2}$, Jie Zheng$^{4,8}$, Wei Wang$^{4,9}$, Zhou Fan$^{4,8}$, Hubiao Niu$^{5}$, Yuan-Yuan Chen$^{6}$, Hao Lu$^{6}$, Xiyan Peng$^{4,8}$, Kai Li$^{7}$, Di-Fu Guo$^{7}$} 

\affil{$^{1}$Department of Astronomy, Yale University, New Haven, CT 06511, USA}
\affil{$^{2}$School of Astronomy and Space Science and Key Laboratory of Modern Astronomy and Astrophysics in Ministry of Education, Nanjing University, Nanjing 210093, China} 
\affil{$^{3}$Mississippi State University, Department of Physics \& Astronomy, Hilbun Hall, Starkville, MS 39762, USA}
\affil{$^{4}$Key Laboratory of Optical Astronomy, National Astronomical Observatories, Chinese Academy of Sciences, Beijing 100012, China} 
\affil{$^{5}$Xinjiang Astronomical Observatory, Chinese Academy of Sciences, Urumqi, Xinjiang 830011, China} 
\affil{$^{6}$Purple Mountain Observatory, Chinese Academy of Sciences, Nanjing 210008, China} 
\affil{$^{7}$Shandong Provincial Key Laboratory of Optical Astronomy and Solar-Terrestrial Environment, Institute of Space Sciences, Shandong University, Weihai 264209, China} 
\affil{$^{8}$University of Chinese Academy of Sciences, Beijing, 100049, China } 
\affil{$^{9}$Chinese Academy of Sciences South America Center for Astronomy, China-Chile Joint Center for Astronomy,
Camino El Observatorio 1515, Las Condes, Santiago, Chile} 
\affil{$^{10}$\textit{51 Pegasi b} Fellow}

\email{song-hu.wang@yale.edu}

\begin{abstract} 

The Kepler-9 system harbors three known transiting planets. The system holds significant interest for several reasons. First, the outer two planets exhibit a period ratio that is close to a 2:1 orbital commensurability, with attendant dynamical consequences. Second, both planets lie in the planetary mass ``desert'' that is generally associated with the rapid gas agglomeration phase of the core accretion process. Third, there exist attractive prospects for accurately measuring both the sky-projected stellar spin-orbit angles as well as the mutual orbital inclination between the planets in the system. Following the original \textit{Kepler} detection announcement in 2010, the initially reported orbital ephemerides for Kepler-9~b and c have degraded significantly, due to the limited time base-line of observations on which the discovery of the system rested. Here, we report new ground-based photometric observations and extensive dynamical modeling of the system. These efforts allow us to photometrically recover the transit of Kepler-9~b, and thereby greatly improve the predictions for upcoming transit mid-times. Accurate ephemerides of this system are important in order to confidently schedule follow-up observations of this system, for both in-transit Doppler measurements as well as for atmospheric transmission spectra taken during transit.
\end{abstract} 

\maketitle
\section{Introduction}

In our own solar system, the major planets all lie within a few degrees of the ecliptic, and the plane containing the planets aligns quite well with the Sun's equator. By contrast, hot Jupiters are frequently observed to have orbital planes that are strikingly misaligned with the equators of their host stars (as reviewed by \citealt{Winn2015}). Spin-orbit angle determinations are made through measurement of the Rossiter-McLaughlin (R-M) effect \citep{Rossiter1924, McLaughlin1924}, a time-variable anomaly in the stellar spectral-line profiles which generates an apparent stellar radial velocity variation during the transit \citep{Queloz2000}. The large observed range in spin-orbit angles is now interpreted as evidence that at least a subset of hot Jupiters were subject to violent dynamical migration at some point in their histories \citep{Winn2015}. This interpretation can be directly tested if the spin-orbit angles can be measured in multi-transiting planetary systems, especially ones hosting planets in mean motion resonance (MMR) configurations. The existence of MMR configurations implies that dissipation in the protoplanetary precusor disk played a key role in sculpting the final planetary orbital configuration, while simultaneously disfavoring a dynamically violent history.

Kepler-9 was the first multiple-planet system discovered using the transit method \citep{Holman2010}. It is also the first definitive TTV system published with an orbital period ratio of outer two planets close to 2:1 commensurability. In addition to suggesting past dissipation generally, such configurations are believed to be the natural consequence of convergent disk migration \citep{Kley2012}. As a consequence, spin-orbit alignment of the planets and the stellar equator is expected. The Kepler-9 system, however, is identified as a candidate misaligned multi-transiting planetary system \citep{Walkowicz2013} based on an assessment with the approximate $v\,{\rm sin}i$ method. Further R-M effect measurement is urgently needed to obtain a definitive answer.

The R-M effect is much more easily measured when transits are deep, and as a practical consequence, R-M observations of multi-transiting planetary systems are hard to make. They usually involve fainter stars and smaller transit depths, and as yet, very few high quality measurements exist (Kepler-89~d, \citealt{Hirano2012, Albrecht2013}; Kepler-25~c, \citealt{Albrecht2013, Benomar2014}; WASP-47~b, \citealt{Sanchis2015}). Both Kepler-9~b and c, however, have very large planet-star size ratios of $R_{\rm b}/R_{*}=0.074$ and $R_{\rm c}/R_{*}=0.076$, respectively \citep{Twicken2016} -- among the largest ratios yet detected in multi-transiting planetary systems. Kepler-9 thus offers a rare opportunity to carry out a high signal-to-noise spin-orbit angle measurement in a multi-transiting planetary system. Successful acquisition will provide a key zeroth-order test of the origin scenarios of spin-orbit misalignments and will potentially help to delineate competing formation paradigms for hot Jupiters.

Moreover, extended transit mid-time measurements will significantly improve the precision of planetary mass determinations, which will aid resolution of the ongoing discrepancy between masses measured through modeling of transit timing variations (TTVs) and masses modeled through accumulation of Doppler radial velocities \citep{Holman2010, Borsato2014, Dreizler2014, Hadden2016}. 

In addition, canonical planet formation theory is challenged by the Kepler-9 planetary masses (Kepler-9~b: $44.17\,{\rm M_\oplus}$, Kepler-9~c: $30.37\,{\rm M_\oplus}$; \citealt{Wang2017c}). In the standard planet formation model \citep{Pollack1996}, the planetary mass should grow rapidly from $30\,{\rm M_\oplus}$ to $100\,{\rm M_\oplus}$ and beyond. It is difficult to arrest run-away gas accretion in the midst of this intermediate range; few planets are known in the ``intermediate mass desert''. It would therefore be of great interest to obtain a high signal-to-noise transmission spectrum to better delineate the atmospheric composition and the planetary structure of this special class of planet.

Any prospective transit-related follow-up studies will require precisely scheduled observations of the transit of Kepler-9~b and c. Such scheduling, however, is not easy. Transit mid-times for Kepler-9~b and c vary by up to $0.78\,$day and $1.78\,$day (Figure~\ref{fig1}), respectively, due to the significant planet-planet gravitational interactions in the system. In this work, we report the successful relocation of Kepler-9~b's transit on UT 2016 Sep. 1, which serves to confirm dynamical modeling of the TTVs derived from the full \textit{Kepler} data set. To provide optimal ephemerides for future observations, we provide the predicted future transit windows of the Kepler-9 system by jointly analyzing the TTVs from both the legacy \textit{Kepler} data as well as our new observations. We also discuss the feasibility of carrying out R-M observations of Kepler-9~b and c, as well as the prospects for directly measuring the mutual inclination between their orbital planes.

This paper adheres to the following organization. In \S 2, we describe the dynamical model that we have employed to predict future transit mid-times for Kepler-9~b and c, based on transits observed during the full 17-quarter \textit{Kepler} data set. In \S 3, we describe our coordinated observations of a Kepler-9~b transit on UT 2016 Sep. 1, along with the data reduction strategies that we employed. In \S 4, we describe the photometric model that we employed to analyze the transit light curve.
In \S 5, the transit mid-time obtained with our light curve is used, together with the complete \textit{Kepler} data set to predict an extensive set of the aperiodic future transits for both Kepler-9~b and c. 
In \S 6, we discuss the prospects for obtaining spin-orbit angle measurements and mutual orbital inclination measurements for the system.
Finally, We summarize the results, and discuss the potential for further investigations in \S 7.

\section{Predicting the Transit Window}

\begin{figure}
\vspace{0cm}\hspace{0cm}
\includegraphics[width=\columnwidth]{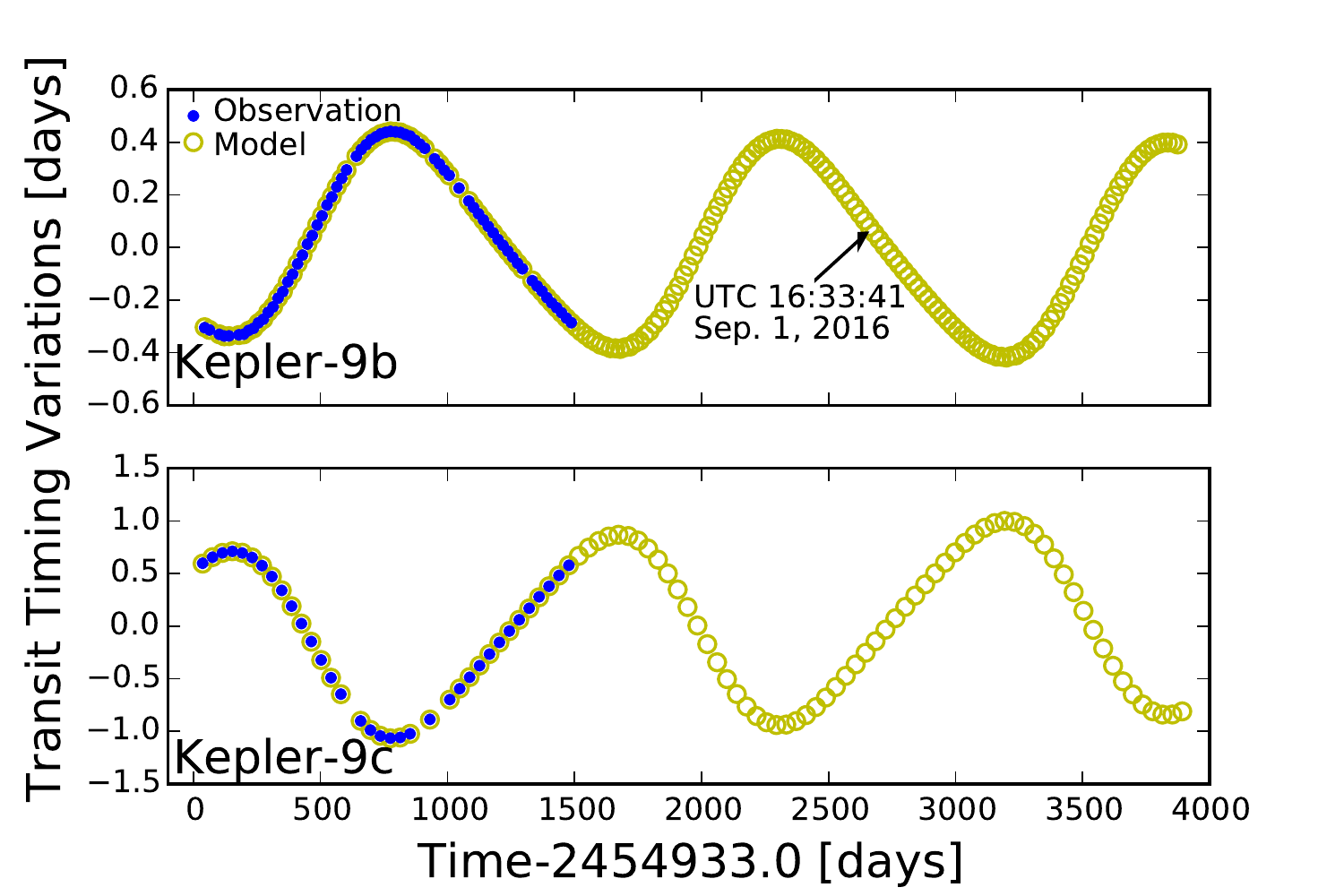}
\centering
\caption{TTVs (blue circles) of Kepler-9~b and c compared to our best fit and prediction (open yellow-green circles) obtained using the full \textit{Kepler} data set. Mutual gravitational interactions between the planets induce TTVs with very high signal-to-noise; the transit mid-times of Kepler-9~b and c vary by up to $0.78\,{\rm day}$ and $1.78\,{\rm day}$, respectively. To facilitate future scheduling of transit-related observations (e.g. the measurement of the R-M effect, or the acquisition of an atmospheric transmission spectrum), we observed Kepler-9 on UT 2016 Sep. 1, when a transit of Kepler-9~b was predicted to occur according to our dynamical model. As indicated with an arrow on the upper panel, the transit was successfully recovered (See Figure~\ref{fig2} for more details). 
\label{fig1}}

\end{figure}

To predict the transit mid-times for Kepler-9~b and c, we carried out a differential evolution Markov chain Monte Carlo-based (DEMCMC, \citealt{Ter2006}) dynamical fitting analysis to infer the orbital parameters of Kepler-9~b and c based on their transit mid-times and uncertainties from the \textit{full} 17-quarters of \textit{Kepler} data \citep{Holczer2016}. Given the mass and the six orbital elements of the planets and the mass of the host star, we calculated the individual transit mid-times of Kepler-9~b and c using a Runga-Kutta-Fehlberg 7(8) N-body code which integrates the full Newtonian equations of motion. The inner planet, Kepler-9~d, is ignored in the model due to its negligible influence on the transit mid-times of Kepler-9~b and c \citep{Dreizler2014}.

\begin{table*}[htbp]
\begin{center}
\begin{threeparttable}[b]
\caption{\bf Dynamical parameters for Kepler-9 system from the best N-body fit to TTVs. The orbital elements are given for the reference epoch $T_{0}=2454900.0 \,{\rm BJD_{TDB}}$.}
\label{tab1}
\begin{tabular}{c|cc|cc}
\tableline\tableline\
&\multicolumn{2}{c|}{\textbf{Kepler data}}&\multicolumn{2}{c}{\textbf{Kepler data + our new data}}\\\hline
\textbf{Parameter}&\textbf{Value}&\textbf{$\sigma$}&\textbf{Value}&\textbf{$\sigma$}\\\hline
$m_{\rm b} [m_{\oplus}]$&43.98&0.47&43.97&0.49\\
$m_{\rm c} [m_{\oplus}]$&30.25&0.33&30.24&0.33\\
$P_{\rm b} [\rm days]$&19.2259&0.000046&19.2259&0.000048\\
$P_{\rm c} [\rm days]$& 39.0153&0.00013&39.0153&0.00013\\
$e_{\rm b}$&0.0637&0.0009&0.0637&0.0009\\
$e_{\rm c}$&0.0680&0.0004&0.0680&0.0003\\
$i_{\rm b} [^\circ]$&89.98&1.01&89.88&1.02\\
$i_{\rm c} [^\circ]$&89.98&0.63&90.01&0.63\\
$\omega_{\rm b} [^\circ]$&357.04&0.44&357.05&0.45\\
$\omega_{\rm c} [^\circ]$&168.65&0.30&168.65&0.31\\
$M_{\rm 0b} [^\circ]$&224.90&0.53&224.88&0.56\\
$M_{\rm 0c} [^\circ]$&253.39&0.29&253.39&0.30\\
$\Delta\Omega [^\circ]$&0.17&1.54&0.43&1.55\\\hline

\end{tabular}
\footnotesize
\end{threeparttable}
\end{center}
\end{table*}

The free parameters considered by the DEMCMC fit are {$P$, $e$, $i$, $\omega$+$M_{0}$, $\omega$-$M_{0}$, $\Delta\Omega$, and $m$}, where $P$ is the orbital period, $e$ is the eccentricity,  $i$ is the orbital inclination, $\omega$ is the argument of periastron, $M_{0}$ is the initial mean anomaly, $\Delta\Omega$ is the difference of the ascending nodes between two planets, and $m$ is the planetary mass. We assume normally distributed priors with median values and uncertainties given in \citet{Dreizler2014}. The central stellar mass is fixed to $1.034 {\rm M_{\odot}}$ \citep{Johnson2017}. We run 40 parallel DEMCMC chains, each with $2\times10^7$ iterations and we save every thousandth set of parameters. The first $1\times10^7$ iterations of each chain are discarded to eliminate burn-in bias. The statistics of the parameters are derived from the final $1\times10^4$ elements of each MCMC assessment. The $\hat{R}$ statistics \citep{Brooks1998} for all parameters were below 1.1 at the conclusion of the calculation. The reported parameters, as detailed in Table \ref{tab1}, are derived  in terms of the medians and standard deviations of the posterior parameter distributions. Our overall best fit  agrees well with the models obtained by \citet{Dreizler2014}, \citet{Borsato2014} , and \citet{Hadden2016}. All four studies, however, report slightly different orbital periods. For further details of Kepler-9 system parameters, we refer the reader to \citet{Wang2017c}.

With the orbital parameters in hand, we can proceed to predict the future transit mid-times for both Kepler-9~b and Kepler-9~c by integrating Newton's equations of motion for the model three-body system. We randomly draw 1000 samples from the converged chains and integrate them independently to determine the future transit mid-times. The best-predicted transit mid-times and the estimated uncertainties (as provided in Table~\ref{tab2} and shown by the open yellow-green circles in Figure~\ref{fig1}), are derived as the median values and standard deviations of the integrated transit mid-time distributions (See Figure~\ref{fig3} for example). The predicted transit mid-times agree with the results from \citet{Dreizler2014} to within 2 min and 3 min for Kepler-9~b and c, respectively.

\section{Observations and Data Reduction}

Using the predicted transit mid-times from our dynamical model, we coordinated observations of a transit of Kepler-9~b on UT 2016 Sep. 1. We secured involvement of eight telescopes at four different observatories in China, all of which are members of the TEMP Network \citep{Wang2017a}. Planned observations from the Xinglong Station (2.16m, 60/90cm Schmidt, 85cm, 80cm, 60cm telescopes), as well as from the Weihai Observatory (1m telescope) were not executed due to the influence of Typhoon Lionrock. The weather at two other observatories, however, was generally cooperative on the transit night. One complete (\S 3.1) and one partial (\S 3.2) transit was obtained.  

\subsection{Nanshan Station, Xinjiang Observatory}

We observed the full transit of Kepler-9~b in a Cousins-$R$ filter using the Nanshan One-meter Wide-field Telescope (Hereafter: NOWT) at Nanshan Station ($\,87^\circ 10' 30''$E,$\,43^\circ 28' 24.66''$N) of the Xinjiang Astronomical Observatory. The NOWT is equipped with a $4{\rm K} \times 4{\rm K}$ CCD that gives a $1.3^{\circ} \times 1.3^{\circ}$ field of view (FOV) and a pixel scale of $1.13''\,{\rm pixel^{-1}}$. For further details of this telescope, see \citet{Liu2014}. 

A $1200 \times 1200$ pixel (approximately $22.5' \times 22.5'$) subframe was used in our observations to shorten the readout time and increase the duty-cycle of the observations. In total, $554$ scientific images were obtained with exposure times varying from $20\,{\rm s}$ to $60\,{\rm s}$ depending on atmospheric conditions. To avoid adversely affecting the measurement of the transit mid-time, the exposure time was not varied during the transit ingress or egress phases. The recorded mid-exposure time in the image header is synchronized with the USNO time server, and was converted to the $\rm{BJD_{TDB}}$ time scale using the techniques of \citet{Eastman2010}; The intrinsic error arising from these conversions is estimated to be less than $1\,{\rm s}$.

All images were debiased and flat-fielded using standard procedures. Aperture photometry was then performed using the DAOPHOT aperture photometry routine \citep{Stetson1987}. The final differential light curve was obtained from weighted ensemble photometry. The choice of photometric comparison stars was made with the goal of minimizing the light-curve scatter. The resulting light curve (yellow-green points) is compared in Figure~\ref{fig2} to the best-fit model.

\subsection{Xuyi Station, Purple Mountain Observatory}

A portion of the transit of Kepler-9~b was also observed through a SDSS-$i$ filter using the 1-m Near Earth Object Survey Telescope (Hereafter: NEOST) located at the Xuyi astronomical station ($\,118^\circ 28'$E,$\,32^\circ 44'$N) of the Purple Mountain Observatory of China. The telescope has a $\rm{10K \times 10K}$ CCD with 16 readout channels. No binning or windowing was used, resulting in a $3^{\circ} \times 3^{\circ}$ FOV, and a pixel scale of $1.029''\,{\rm \, pixel^{-1}}$. 

We obtained 251 images at an observing cadence of $\sim77\,{\rm s}$ comprising a $60\,\rm{s}$ exposure time and a $17\,{\rm s}$ readout/reset time between exposures. Due to clouds, the observations were interrupted for about $\,{\rm 30\,min}$ during the ingress phase, and the observations had to be stopped during the transit because of the elevation limits of the telescope.

The time recording approach and data reduction strategy were identical to the method used above (\S 3.1). The resulting light curve (blue points) is compared in Figure~\ref{fig2} to the best-fit model.

\section{Light Curve Analysis}

In order to accurately measure the transit mid-time of our new transit observation of Kepler-9~b, we performed a joint fit to the NOWT and NEOST light curves. The photometric modeling was carried out using JKTEBOP code \citep{Southworth2008}, which employs Levenberg--Marquardt minimization to find the best fit, and a residual-permutation algorithm to determine the error estimates for the derived parameters.

Due to the limited quality of our new ground-based light curve of the Kepler-9~b transit, a refinement of the system parameters was not a goal in this work. Moreover, while there are substantial variations in the transit mid-times, none of the other transit parameters for Kepler-9~b showed significant deviation over the 4-year duration of \textit{Kepler} observations \citep{Holman2010, Dreizler2014}. Therefore, we estimated the mid-time of our new Kepler-9~b transit by allowing \textit{only} the transit mid-time, $T_{\rm c}$, as well as the light-curve specific base-line flux, $F_{0}$, to float, while holding the remaining parameters fixed. We fixed the basic transit parameters -- the planet-to-star radius ratio, $R_{\rm P}/R_{*}$, the scaled semimajor axis, ${a}/({R_{*}+R_{\rm P}})$, and the orbital inclination, $i$, -- at the best-fit values derived from the high-precision \textit{Kepler} transit light curves \citep{Twicken2016}. We held the orbital period, $P$, and the two quantities $e\cdot{\rm cos}\,\omega$ and $e\cdot{\rm sin}\,\omega$ relating the eccentricity, $e$, and the argument of periastron, $\omega$, (which are typically poorly constrained by a single transit observation) fixed to the values determined from the TTV analysis described in \S 2. A limb-darkening law containing both a linear and a quadratic term was adopted, with coefficients fixed to the tabulated values $\mu_{1,R}=0.37$ and $\mu_{2,R}=0.28$, using the spectroscopic stellar parameters -- effective temperature, $T_{\rm eff}=5779\,{\rm K}$, metallicity, $\rm{[Fe/H]}=0.12$, and surface gravity, ${\rm log}\,g=4.491$, -- from \citet{Huber2014}.

\begin{figure}
\vspace{0cm}\hspace{0cm}
\centering
\includegraphics[width=\columnwidth]{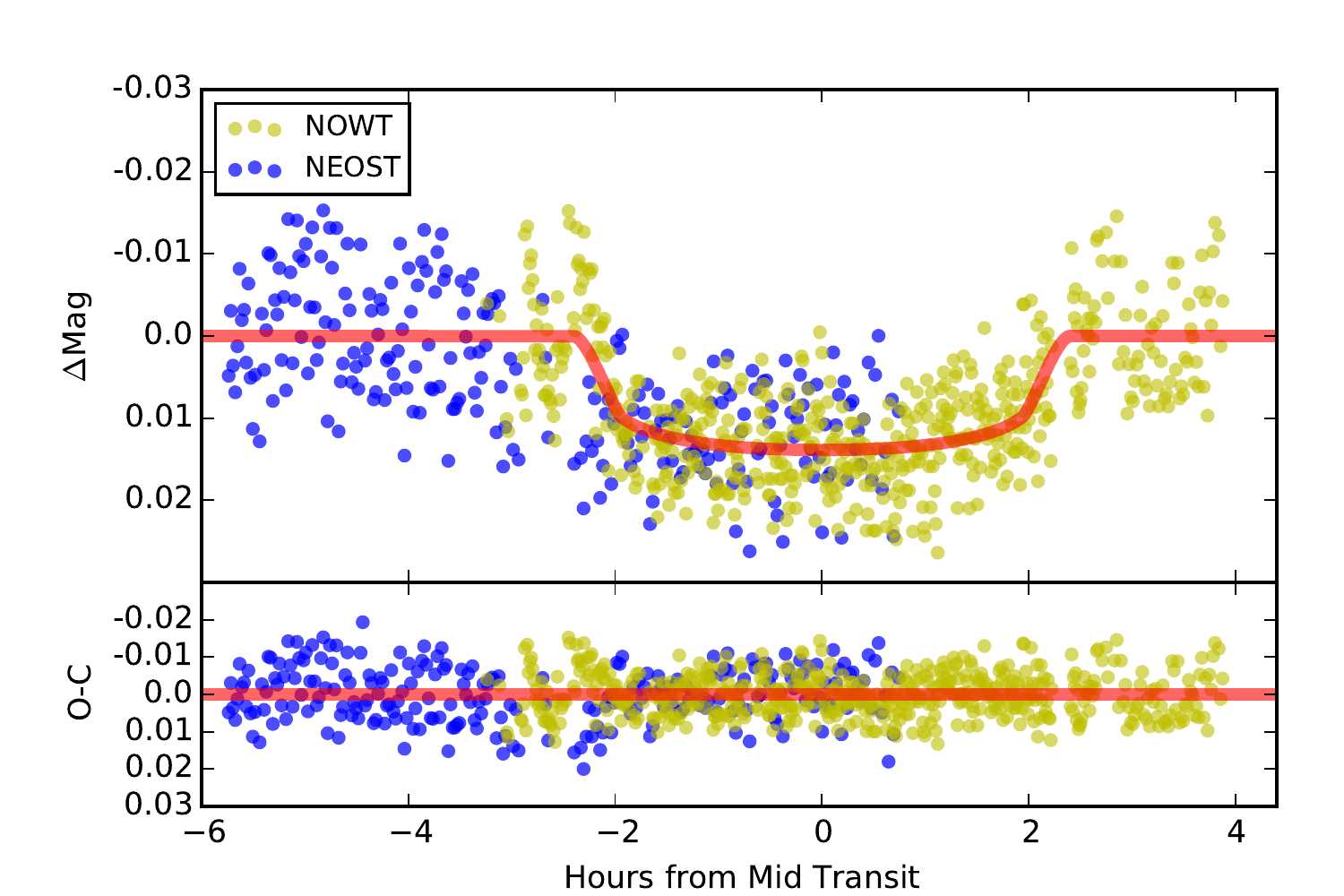}
\caption{Transit light curves for Kepler-9~b, acquired on UT 2016 Sep. 1 with the NOWT (yellow-green points) and NEOST (blue points) telescopes. The solid red line shows the best-fit model. The residuals of light curve from the best-fit model are plotted in the lower panel of the figure.
\label{fig2}}
\end{figure}

\begin{figure}
\vspace{0cm}\hspace{0cm}
\centering
\includegraphics[width=\columnwidth]{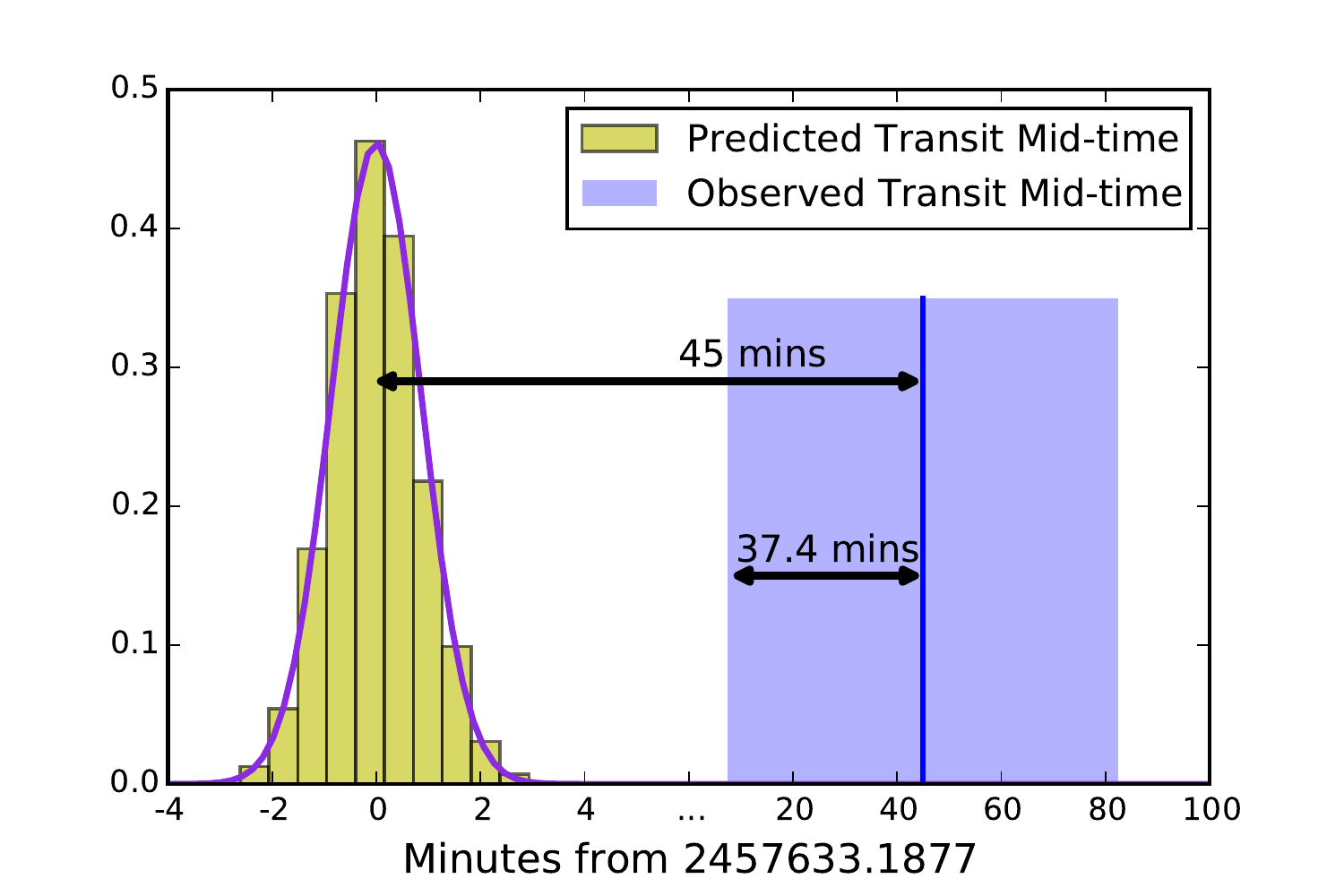}
\caption{Predicted mid-time vs. Observed mid-time for Kepler-9~b's transit on UT 2016 Sep. 1. 
It is unclear how such a $\sim45\,$min ($1.2\,\sigma$) discrepancy between prediction and observation could arise, although one possible source would be the secular TTV effect from an additional unknown companion or companions in the system.
Further photometric and velocimetric observation of Kepler-9 system are recommended to help understand this discrepancy.
\label{fig3}}
\end{figure}

The best fit to the new transit light curves for Kepler-9~b is plotted in Figure~\ref{fig2}.
The measured transit mid-time is $T_{\rm}=2457633.219 \pm 0.026$, which is in good agreement with the transit mid-time derived using only the complete transit obtained with NOWT. (NEOST's light curve alone, without sampling of the ingress and egress cannot be used for transit mid-time estimation). Our new observations robustly reveal a transit event for Kepler-9~b, but as shown in Figure~\ref{fig3}, the transit event occurs $45\,$min ($1.2\,\sigma$) later than predicted. We cannot claim a large significance for this discrepancy between the predicted and the observed transit mid-times for Kepler-9~b's UT 2016 Sep. 1 transit, as substantial systematic errors may have affected the determination of transit mid-time. There may be, however, some possible physical reasons for the 45-minute discrepancy. Since our new observation occurred three years after the last \textit{Kepler} observation, secular TTV drifts arising from an unknown stellar or planetary companion(s), may have contributed to the deviation. A similar conclusion was also reached by \citet{Dreizler2014} based on the substantial RMS scatter in the Doppler velocity residuals of Kepler-9 system, which are consistent with the possibility that is harboring additional companions. Unfortunately, by adding only one new transit, we are unable to shed definitive light. The collection of further photometry and velocimetry would appear quite well advised.

\section{Improved Transit Predictions}

Since we have successfully observed the transit of Kepler-9~b on UT 2016 Sep. 1, and obtained the transit mid-time by modeling the light curve, we refit the orbital parameters of Kepler-9~b and c based on transit mid-times from both \textit{Kepler} data and our new data using the approach described in \S 2. With only one more transit observed, the rederived parameters remained almost the same, aside from a small increase in uncertainties as shown in Table~\ref{tab1}. The corresponding future transit mid-times also show little variation with those calculated with \textit{Kepler} data only  (Table~\ref{tab2}).

\begin{table*}[htbp]
\begin{center}
\caption{\bf Prediction of aperiodic future transit mid-times(day) for Kepler-9~b and c from our best N-body fit to TTVs. The time is given as $\rm{BJD_{TDB}}-2454900.0$.} 
\label{tab2}
\begin{tabular}{c|c|c|c}
\tableline\tableline
\multicolumn{2}{c|}{\textbf{Kepler data}}&\multicolumn{2}{c}{\textbf{Kepler data + our new data}}\\\hline
\multicolumn{1}{c|}{Kepler-9~b}&\multicolumn{1}{c|}{Kepler-9~c}&\multicolumn{1}{c|}{Kepler-9~b}&\multicolumn{1}{c|}{Kepler-9~c}\\\hline
       $2733.1877^{\pm0.0006}$	&	       $2718.1377^{\pm0.0017}$	&	       $2733.1877^{\pm0.0006}$	&	       $2718.1377^{\pm0.0017}$	\\
       $2752.4069^{\pm0.0006}$	&	       $2757.2123^{\pm0.0017}$	&	       $2752.4069^{\pm0.0006}$	&	       $2757.2123^{\pm0.0017}$	\\
       $2771.6269^{\pm0.0006}$	&	       $2796.2862^{\pm0.0017}$	&	       $2771.6269^{\pm0.0006}$	&	       $2796.2862^{\pm0.0017}$	\\
       $2790.8465^{\pm0.0006}$	&	       $2835.3593^{\pm0.0017}$	&	       $2790.8464^{\pm0.0006}$	&	       $2835.3593^{\pm0.0018}$	\\
       $2810.0672^{\pm0.0006}$	&	       $2874.4314^{\pm0.0018}$	&	       $2810.0672^{\pm0.0006}$	&	       $2874.4314^{\pm0.0018}$	\\
       $2829.2872^{\pm0.0006}$	&	       $2913.5022^{\pm0.0018}$	&	       $2829.2872^{\pm0.0007}$	&	       $2913.5022^{\pm0.0019}$	\\
       $2848.5087^{\pm0.0006}$	&	        $2952.571^{\pm0.0018}$	&	       $2848.5087^{\pm0.0007}$	&	        $2952.571^{\pm0.0019}$	\\
       $2867.7291^{\pm0.0007}$	&	       $2991.6363^{\pm0.0019}$	&	       $2867.7291^{\pm0.0007}$	&	       $2991.6364^{\pm0.0019}$	\\
       $2886.9513^{\pm0.0007}$	&	       $3030.6962^{\pm0.0018}$	&	       $2886.9513^{\pm0.0007}$	&	       $3030.6962^{\pm0.0019}$	\\
       $2906.1724^{\pm0.0007}$	&	       $3069.7477^{\pm0.0018}$	&	       $2906.1724^{\pm0.0007}$	&	       $3069.7478^{\pm0.0019}$	\\
       $2925.3952^{\pm0.0007}$	&	       $3108.7875^{\pm0.0016}$	&	       $2925.3952^{\pm0.0007}$	&	       $3108.7876^{\pm0.0018}$	\\
       $2944.6172^{\pm0.0007}$	&	       $3147.8112^{\pm0.0015}$	&	       $2944.6172^{\pm0.0007}$	&	       $3147.8113^{\pm0.0017}$	\\
       $2963.8408^{\pm0.0007}$	&	       $3186.8146^{\pm0.0015}$	&	       $2963.8408^{\pm0.0008}$	&	       $3186.8147^{\pm0.0017}$	\\
        $2983.064^{\pm0.0008}$	&	       $3225.7932^{\pm0.0017}$	&	        $2983.064^{\pm0.0008}$	&	       $3225.7934^{\pm0.0019}$	\\
       $3002.2885^{\pm0.0008}$	&	       $3264.7436^{\pm0.0022}$	&	       $3002.2885^{\pm0.0008}$	&	       $3264.7438^{\pm0.0023}$	\\
       $3021.5133^{\pm0.0008}$	&	       $3303.6635^{\pm0.0027}$	&	       $3021.5133^{\pm0.0008}$	&	       $3303.6637^{\pm0.0028}$	\\
       $3040.7391^{\pm0.0008}$	&	       $3342.5525^{\pm0.0033}$	&	       $3040.7391^{\pm0.0009}$	&	       $3342.5527^{\pm0.0033}$	\\
       $3059.9661^{\pm0.0009}$	&	       $3381.4122^{\pm0.0038}$	&	       $3059.9661^{\pm0.0009}$	&	       $3381.4125^{\pm0.0038}$	\\
       $3079.1937^{\pm0.0009}$	&	       $3420.2461^{\pm0.0042}$	&	       $3079.1936^{\pm0.0009}$	&	       $3420.2465^{\pm0.0042}$	\\
       $3098.4237^{\pm0.0009}$	&	       $3459.0591^{\pm0.0044}$	&	        $3098.4236^{\pm0.001}$	&	       $3459.0595^{\pm0.0044}$	\\
        $3117.6535^{\pm0.001}$	&	       $3497.8569^{\pm0.0045}$	&	        $3117.6535^{\pm0.001}$	&	       $3497.8573^{\pm0.0046}$	\\
       $3136.8874^{\pm0.0011}$	&	       $3536.6458^{\pm0.0045}$	&	       $3136.8873^{\pm0.0011}$	&	       $3536.6462^{\pm0.0046}$	\\
       $3156.1203^{\pm0.0011}$	&	       $3575.4325^{\pm0.0044}$	&	       $3156.1202^{\pm0.0012}$	&	        $3575.433^{\pm0.0044}$	\\
       $3175.3589^{\pm0.0013}$	&	       $3614.2238^{\pm0.0042}$	&	       $3175.3589^{\pm0.0013}$	&	       $3614.2242^{\pm0.0043}$	\\
       $3194.5957^{\pm0.0013}$	&	       $3653.0262^{\pm0.0039}$	&	       $3194.5956^{\pm0.0013}$	&	        $3653.0266^{\pm0.004}$	\\
         $3213.84^{\pm0.0015}$	&	       $3691.8462^{\pm0.0036}$	&	       $3213.8399^{\pm0.0015}$	&	       $3691.8466^{\pm0.0037}$	\\
       $3233.0814^{\pm0.0016}$	&	       $3730.6898^{\pm0.0032}$	&	       $3233.0813^{\pm0.0016}$	&	       $3730.6902^{\pm0.0034}$	\\
       $3252.3321^{\pm0.0018}$	&	       $3769.5617^{\pm0.0029}$	&	        $3252.332^{\pm0.0018}$	&	       $3769.5621^{\pm0.0031}$	\\
       $3271.5788^{\pm0.0019}$	&	       $3808.4654^{\pm0.0026}$	&	       $3271.5787^{\pm0.0019}$	&	       $3808.4658^{\pm0.0029}$	\\
       $3290.8364^{\pm0.0021}$	&	        $3847.402^{\pm0.0025}$	&	       $3290.8363^{\pm0.0021}$	&	       $3847.4024^{\pm0.0028}$	\\
       $3310.0887^{\pm0.0022}$	&	       $3886.3706^{\pm0.0026}$	&	       $3310.0886^{\pm0.0022}$	&	       $3886.3709^{\pm0.0029}$	\\
       $3329.3532^{\pm0.0024}$	&	       $3925.3682^{\pm0.0027}$	&	       $3329.3531^{\pm0.0024}$	&	       $3925.3684^{\pm0.0031}$	\\
       $3348.6114^{\pm0.0025}$	&	        $3964.3903^{\pm0.003}$	&	       $3348.6113^{\pm0.0025}$	&	       $3964.3905^{\pm0.0033}$	\\
       $3367.8824^{\pm0.0027}$	&		&	       $3367.8822^{\pm0.0026}$	&		\\
       $3387.1461^{\pm0.0028}$	&		&	        $3387.146^{\pm0.0027}$	&		\\
       $3406.4227^{\pm0.0029}$	&		&	       $3406.4226^{\pm0.0028}$	&		\\
        $3425.6915^{\pm0.003}$	&		&	       $3425.6913^{\pm0.0029}$	&		\\
       $3444.9726^{\pm0.0031}$	&		&	        $3444.9724^{\pm0.003}$	&		\\
       $3464.2454^{\pm0.0031}$	&		&	        $3464.2452^{\pm0.003}$	&		\\
       $3483.5297^{\pm0.0032}$	&		&	       $3483.5295^{\pm0.0031}$	&		\\
       $3502.8055^{\pm0.0032}$	&		&	       $3502.8053^{\pm0.0031}$	&		\\
       $3522.0915^{\pm0.0032}$	&		&	       $3522.0913^{\pm0.0032}$	&		\\
       $3541.3691^{\pm0.0032}$	&		&	       $3541.3689^{\pm0.0031}$	&		\\
       $3560.6554^{\pm0.0032}$	&		&	       $3560.6552^{\pm0.0031}$	&		\\
       $3579.9334^{\pm0.0032}$	&		&	       $3579.9332^{\pm0.0031}$	&		\\
       $3599.2184^{\pm0.0031}$	&		&	       $3599.2182^{\pm0.0031}$	&		\\
       $3618.4956^{\pm0.0031}$	&		&	        $3618.4954^{\pm0.003}$	&		\\
        $3637.7779^{\pm0.003}$	&		&	        $3637.7776^{\pm0.003}$	&		\\
        $3657.0529^{\pm0.003}$	&		&	       $3657.0527^{\pm0.0029}$	&		\\
       $3676.3309^{\pm0.0029}$	&		&	       $3676.3306^{\pm0.0029}$	&		\\
       $3695.6024^{\pm0.0028}$	&		&	       $3695.6022^{\pm0.0028}$	&		\\
       $3714.8748^{\pm0.0027}$	&		&	       $3714.8746^{\pm0.0027}$	&		\\
       $3734.1418^{\pm0.0026}$	&		&	       $3734.1416^{\pm0.0026}$	&		\\
       $3753.4075^{\pm0.0026}$	&		&	       $3753.4072^{\pm0.0026}$	&		\\
       $3772.6688^{\pm0.0025}$	&		&	       $3772.6686^{\pm0.0025}$	&		\\
        $3791.927^{\pm0.0024}$	&		&	       $3791.9268^{\pm0.0024}$	&		\\
       $3811.1821^{\pm0.0023}$	&		&	       $3811.1819^{\pm0.0023}$	&		\\
       $3830.4326^{\pm0.0022}$	&		&	       $3830.4324^{\pm0.0022}$	&		\\
       $3849.6811^{\pm0.0021}$	&		&	       $3849.6809^{\pm0.0022}$	&		\\
         $3868.924^{\pm0.002}$	&		&	       $3868.9238^{\pm0.0021}$	&		\\
        $3888.1661^{\pm0.002}$	&		&	        $3888.1659^{\pm0.002}$	&		\\
       $3907.4021^{\pm0.0019}$	&		&	         $3907.402^{\pm0.002}$	&		\\
       $3926.6383^{\pm0.0018}$	&		&	       $3926.6382^{\pm0.0019}$	&		\\
       $3945.8686^{\pm0.0018}$	&		&	       $3945.8684^{\pm0.0019}$	&		\\
       $3965.0996^{\pm0.0017}$	&		&	       $3965.0995^{\pm0.0018}$	&		\\
       $3984.3252^{\pm0.0017}$	&		&	       $3984.3251^{\pm0.0018}$	&		\\\hline

\end{tabular}
\end{center}
\end{table*}

\section{Stellar Obliquity and Mutual Orbital Inclination}

\begin{figure}
\includegraphics[width=\columnwidth]{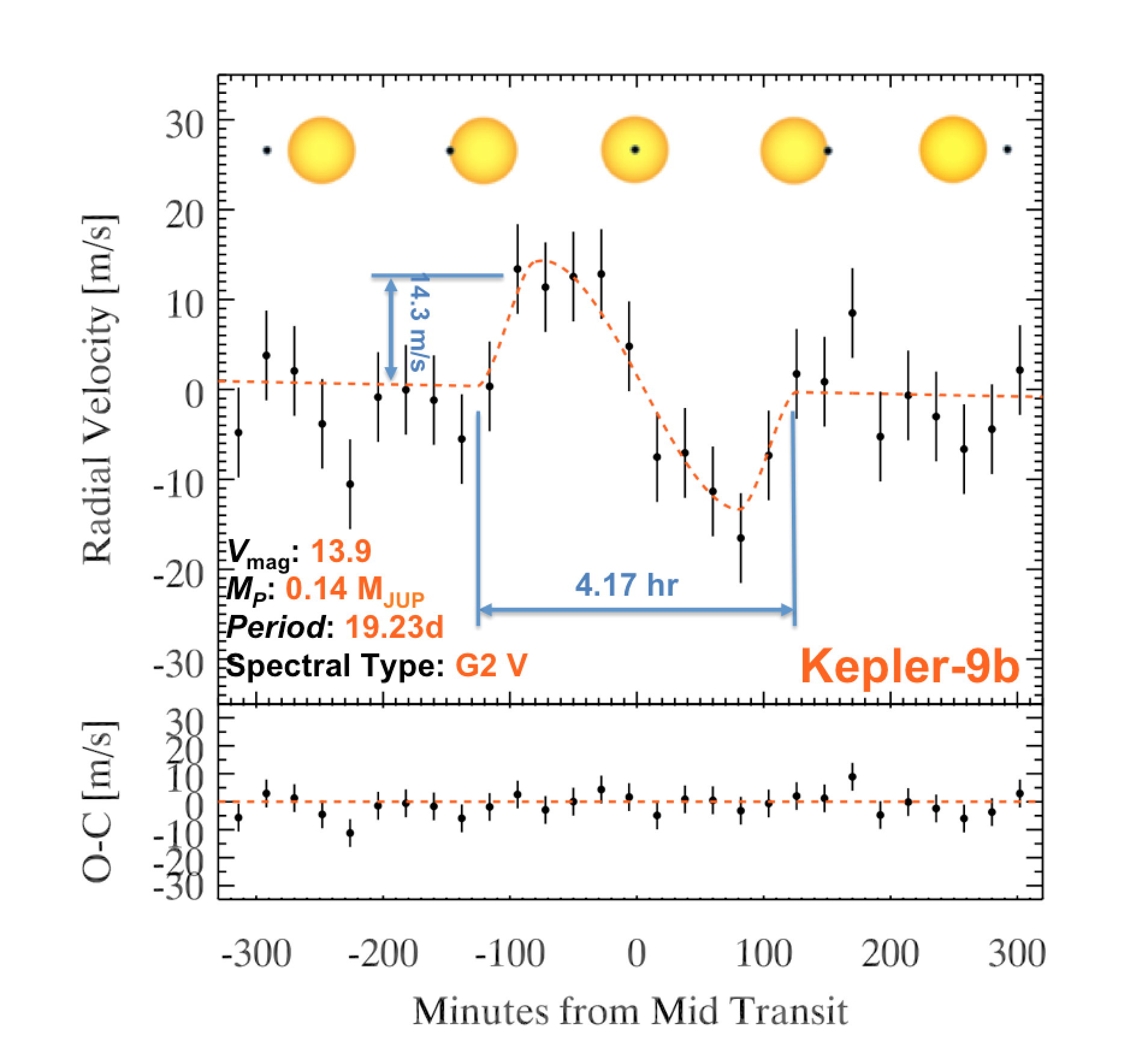}
\caption{Synthetic Keck/HIRESr radial velocities of the Kepler-9~b transit with an assumed $v\,{\rm sin}\,i=2200\,{\rm m\,s^{-1}}$, $\lambda=0$, and $20\,{\rm min}$ exposure ($\sigma_{\rm rv}=5\,{\rm m\,s^{-1}}$). The best-fit R-M effect is over-plotted as dashed orange line. The bottom panel shows the residuals of the best fit. The spin-orbit angle of Kepler-9~b can be measured to $\pm 5.3^{\circ}$.}
\label{fig4}
\end{figure}

Although R-M observations of multi-transiting planetary systems may play a significant role in shaping the observed properties of exoplanets, and distinguishing the competing formation scenarios for hot Jupiters, they are hard to make. The R-M effect is most easily observed when transits are deep and when the target star is bright. Multi-transiting planetary systems, however, usually involve smaller transit depths and fainter parent stars.  

The Kepler-9 system, however, is an exception. It has a very large planet-star size ratio -- among the largest ratios yet detected in multi-transiting planetary systems, and it is a proven target for obtaining high-precision radial velocities with $2.5\,{\rm m\,s^{-1}}$ error using $45\,{\rm min}$ exposures with HIRESr on the Keck I telescope \citep{Holman2010}. Indeed Kepler-9 may offer a rare to measure the spin-orbit angle of a multi-transiting planetary system. We have therefore made a detailed examination of the feasibility of carrying out R-M observations of Kepler-9~b and c using Keck/HIRESr.

\begin{figure}
\includegraphics[width=\columnwidth]{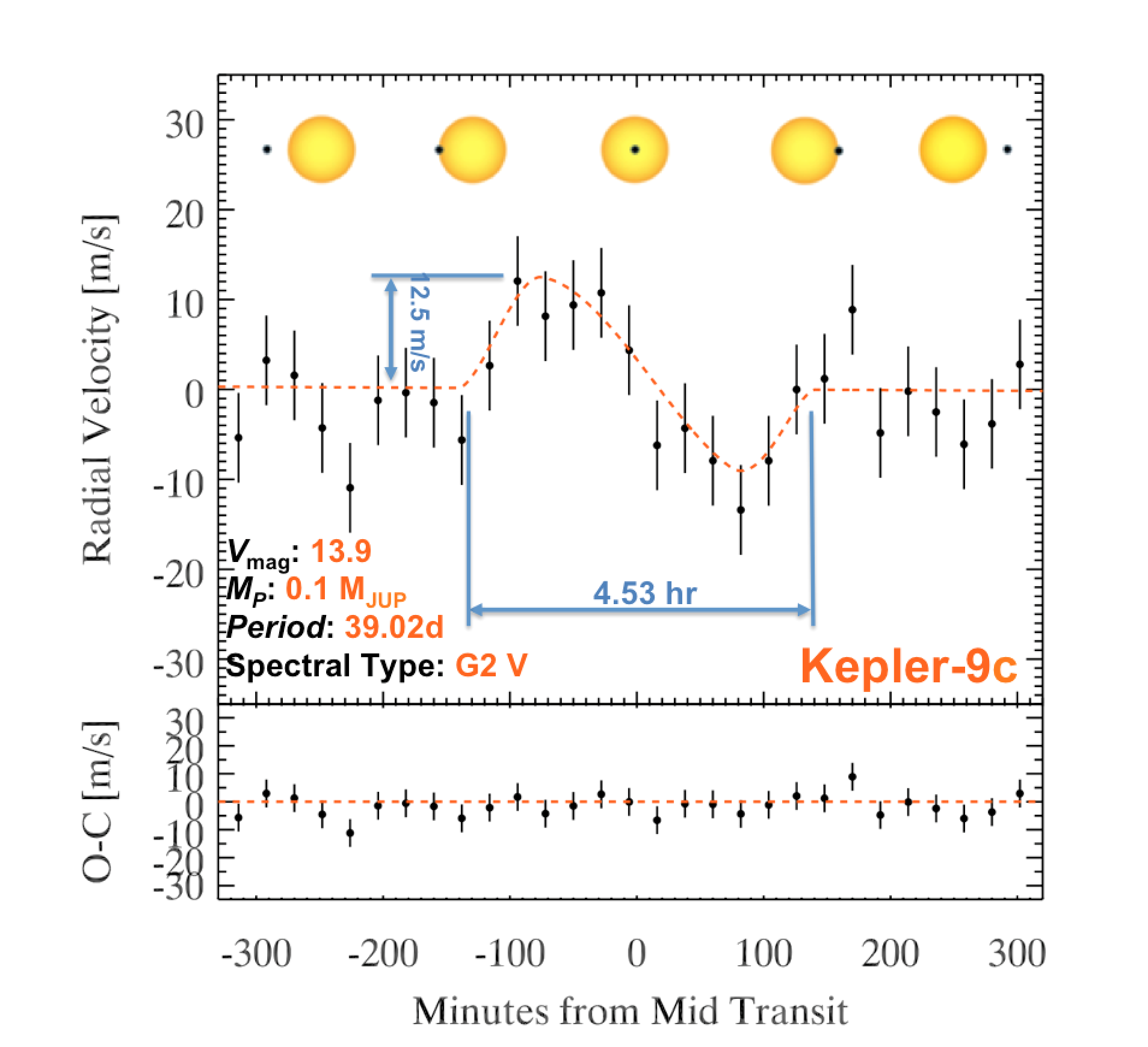}
\caption{Same as Figure~\ref{fig4} but for Kepler-9~c. The spin-orbit angle of Kepler-9~c can be measured to $\pm 6.3^{\circ}$. The measurement of spin-orbit angles for two different planets in the same system, like Kepler-9, will provide us an \textit{unique} chance to directly measure their mutual orbital inclination (See Equation~\ref{eq1} for more details).}
\label{fig5}
\end{figure}

This was done by first generating synthetic radial velocities of the R-M effect with added Gaussian noise, assuming both Kepler-9~b and c are in spin-orbit alignments. To span the transit with at least 10 observations, we set the length of each exposure to $20\,{\rm mins}$. This provides us with 11 and 12 radial velocities during the Kepler-9~b and c transits, respectively, with uncertainties of $5\,{\rm m\,s^{-1}}$ \citep{Burt2015}, which is consistent with the radial velocity precision achieved by \citet{Holman2010} with Keck/HIRESr. The R-M effect induced velocity anomaly is modeled using the analytical approach of \citet{Hirano2010} and is discussed in detail in \citet{Addison2013}. The parameters used in the R-M effect modeling include: the spin-orbit angle, $\lambda$, the stellar rotation velocity, $v\,{\rm sin}\,i$, the planet-to-star radius ratio, $R_{\rm P}/R_*$, the orbital inclination, $i$, the orbital period, $P$, the planet-to-star mass ratio, $M_{\rm P}/M_*$, the orbital eccentricity, $e$, the argument of periastron, $\omega$, and two adopted quadratic limb-darkening coefficients, $\mu_1$, and $\mu_2$. $\lambda$ is set to 0, the spin of the star is modeled with $v\,{\rm sin}\,i=2200\,{\rm m\,s^{-1}}$ \citep{Buchhave2012}, and the other parameters are set to the values given in \citet{Twicken2016} and Table~\ref{tab1} (Parameters derived from \textit{Kepler} data+ our new data). The predicted R-M anomalies for Kepler-9~b and c are shown in Figures~\ref{fig4} and \ref{fig5} with half-amplitudes of $14.3\,{\rm m\,s^{-1}}$ and $12.5\,{\rm m\,s^{-1}}$, respectively.

After generating the synthetic radial velocities, we used the Exoplanetary Orbital and Simulation and Analysis Model (ExOSAM; see \citealt{Addison2016}) to fit the generated R-M observations to recover the spin-orbit angle and probe the level of uncertainty that can be measured. Gaussian distributions are assumed on the priors for Kepler-9~b and c, based on their best-fit parameter values and $1\,{\sigma}$ uncertainties given in \citet{Twicken2016} and Table~\ref{tab1} (Parameters derived from \textit{Kepler} data augmented by our new data). To determine the best-fit $\lambda$, we imposed a Gaussian prior on $v\,{\rm sin}\,i$ of $2200\pm1000 \,{\rm m\,s^{-1}}$ \citep{Buchhave2012}.

We generated $20,000$ accepted MCMC iterations with an acceptance rate of $25\%$. This ensured good convergence and thorough mixing of the Markov chains. The mean and standard deviation of the MCMC chains were then used to compute $\lambda$ and $\sigma_{\lambda}$, resulting in $-2.5\pm5.3\,{\rm deg}$ and $-7.5\pm6.3\,{\rm deg}$ for Kepler-9~b and c, respectively. The best-fit models to the simulated radial velocities are over-plotted in Figures~\ref{fig4} and \ref{fig5}. Our simulations strongly suggest that the spin-orbital angles of Kepler-9~b and c can be successfully measured with uncertainties less than $10\,{\rm deg}$ using Keck/HIRESr.

The assessment of the degree of projected spin-orbit angles for two different planets in a same system such as Kepler-9, will provide us a unique chance to directly measure mutual orbital inclination $\delta$ through: 
\begin{align}
\label{eq1}
\cos{\delta}=\cos{i_{\rm b}}\cos{i_{\rm c}}+\sin{i_{\rm b}}\sin{i_{\rm c}}\cos(\lambda_{\rm b}-\lambda_{\rm c}),
\end{align}
where $i_{\rm b,c}$ and $\lambda_{\rm b,c}$ are the orbital inclinations and the sky-projected spin-orbit angles for two planets.

In our own solar system, the major planets all lie within a few degrees of the ecliptic. This is compelling evidence that the Solar System planets originated within a flat rotating disk. It is still unclear, however, whether other multiple-planet systems, especially those with significantly massive planets, also generally adhere to this organizational principle, because directly measuring mutual orbital inclination between the planets is challenging \citep{Winn2015}.  Kepler-9 is a unique target to carry out the mutual orbital inclination measurement through measuring the spin-orbit angles for both Kepler-9~b and c, which will provides us the fundamental architectural data that will answer this critical question.

\section{Summary \& Conclusion}
The Kepler-9 system imparts significant historical interest as a consequence of exhibiting the first observed exoplanetary TTVs. These inconstant eclipses confirmed the promise of the forward-looking analyses of \citet{Agol2005} and \citet{Holman2005}, and set the stage for many insights to emerge from the study of TTVs, culminating with mass estimates for the Earth-sized planets in the TRAPPIST-1 system by \citet{Gillon2017} and \citet{Wang2017b}. In this paper, using both the full complement of \textit{Kepler} photometry and a ground-based photometric recovery of the transit of Kepler-9~b, we have substantially improved the transit ephemerides for both planets b and c. These predictions will permit the accurate scheduling of follow-up measurements with large space-based or ground-based telescopes. \
\\
\\
\textbf{Acknowledgments}  \\ 

We are thankful to the anonymous referee for providing helpful comments that greatly improved the manuscript.

S.W. thank the Heising-Simons Foundation for their generous support.

This research is supported by Nanshan 1m telescope of Xinjiang Astronomical Observatory; the CAS ``Light of West China'' program (2015-XBQN-A-02); the National Basic Research Program of China (Nos. 2014CB845704, and 2013CB834902); the National Natural Science Foundation of China under grant Nos. 11503009, 11333002, 11373033, 11373003, 11633009, 11233004, 11403107, 11503090, 11273067, 
11433005, 11673027, and U1631102; the Minor Planet Foundation of Purple Mountain Observatory; the Natural Science Foundation of Jiangsu Province (BK20141045); the Chinese Academy of Sciences (CAS), through a grant to the CAS South America Center for Astronomy (CASSACA) in Santiago, Chile.

We thank Xiaojia Zhang for her help improving Figure~\ref{fig4} and \ref{fig5}.

\end{document}